\documentclass[prl,superscriptaddress,twocolumn,letterpaper,amssymb,amsmath,floatfix]{revtex4-1}
\pdfoutput=1
\usepackage{gensymb}
\usepackage{color}
\usepackage{graphicx}
\usepackage{mathrsfs}
\usepackage{epstopdf}

\begin{document}

\title{ Fractionalized spin-wave continuum in spin liquid states on the kagome
lattice }

\author{Jia-Wei Mei} \affiliation{Perimeter Institute for Theoretical Physics,
Waterloo, Ontario, N2L 2Y5 Canada} \author{Xiao-Gang Wen}
\affiliation{Department of Physics, Massachusetts Institute of Technology,
Cambridge, Massachusetts 02139, USA} \affiliation{Perimeter Institute for
Theoretical Physics, Waterloo, Ontario, N2L 2Y5 Canada} \date{\today}
\begin{abstract}

Motivated by spin-wave continuum (SWC) observed in recent neutron scattering
experiments in Herbertsmithite, we use Gutzwiller-projected wave functions to
study dynamic spin structure factor $S(\mathbf{q},\omega)$ of spin liquid
states on the kagome lattice. Spin-1 excited states in spin liquids are
represented by Gutzwiller-projected two-spinon excited wave functions. We
investigate three different spin liquid candidates, spinon Fermi-surface spin
liquid (FSL), Dirac spin liquid (DSL) and random-flux spin liquid (RSL). FSL
and RSL have low energy peaks in $S(\mathbf{q},\omega)$ at $K$ points in the
extended magnetic Brillouin zone, in contrast to experiments where low energy
peaks are found at $M$ points. There is no obviuos contradiction between DSL
and neutron scattering measurements. Besides a fractionalized spin ({\it i.e.}
spin-1/2), spinons in DSL carry a fractionalized crystal momentum which is
potentially detectable in SWC in the neutron scattering measurements.

\end{abstract} \maketitle

Quantum spin liquid states has been catching more and more attention in
condensed matter physics\cite{Anderson1987,Lee2008,Balents2010}. They are new
states of matters that are beyond the description of Landau's symmetry breaking
theory of conventional ordered phases\cite{Wen2004}. Spin degrees of freedom in
quantum spin liquids are not frozen at zero temperatures, but highly entangled
with one another over long ranges. The symmetries in long-range entangled
many-body systems can be fractionalized\cite{Laughlin1983,Wen2002}. Quantum
spin liquids allow deconfined spinon excitations which carry a fractional spin
({\it i.e.} spin-1/2) and  give rise to spin-wave continuum (SWC) through a
pair of spinon particle-hole
excitations\cite{Read1989,Kivelson1989,Read1991,Wen1991}.  In some spin liquid
states, spinons carry ``fractional crystal momenta''. As a result, the momentum
resolved density of states for spin-1 excitation continuum ({\it i.e.}
two-spinon excitations) has a period smaller than the elementary Brillouin zone
(BZ) which is potentially detectable in neutron scattering
measurements\cite{Wen2002,Wen2004}.

Herbertsmithite [ZnCu$_3$(OH)$_6$Cl$_2$], a layered spin-1/2 kagome lattice
antiferromagnet, is a promising compound for an experimental realization of
spin liquid
states\cite{Shores2005,Helton2007,Mendels2007,Zorko2008,Imai2008,Vries2009,Imai2011,Jeong2011,Han2012,Han2012a}.
Recent inelastic neutron scattering on single crystals of
Herbertsmithite\cite{Han2012} has detected a diffuse low energy SWC over a
large energy and momentum regions. Nearest-neighbor kagome antiferromagnetic
Heisenberg model (KAFHM) has been suggested for  spin-liquid  physics  observed
in Herbertsmithite. Many different ground states have been proposed for
KAFHM\cite{Marston1991,Sachdev1992,Mila1998,Hastings2000,Nikolic2003,Budnik2004,Wang2006,Ran2007,Singh2008,Hermele2008,Jiang2008,Evenbly2010,Poilblanc2010,Lu2011,Iqbal2013}.
Recent density-matrix-renormalization-group
calculations\cite{Yan2011,Jiang2012,Depenbrock2012} support a $\mathbb{Z}_2$
gapped spin liquid ground state and indicate the kagome ground state is
proximate to a critical state. The low-energy spin excitations are also studied
for these proposed candidate ground
states\cite{Budnik2004,Singh2008,Messio2010,Hao2010,Dodds2013,Rousochatzakis2014,Punk2014}
and in the exact diagonalization\cite{Laeuchli2009}.

In this letter, we will compute dynamic spin structure factor
$S(\mathbf{q},\omega)$ for spin liquid states on the kagome lattice. The ground
state for a spin liquid is described by the Gutzwiller-projected wave function
(GPWF) by projecting out double occupancy components in the mean field ground
state\cite{Anderson1987,Gros1989}. Similarly, GPWFs for spin-1 excited states
are constructed by applying Gutzwiller projection onto spinon-antispinon
excited wave functions\cite{Li2010,DallaPiazza2015}. As well as equal-time spin
factor $S(\mathbf{q})$ in the ground-state GPWF, we use Monte Carlo method to
calculate the projected Hamiltonian system $\{\mathbb{H},\mathbb{O}\}$ where
$\mathbb{H}$ and $\mathbb{O}$ are the Hamiltonian matrix and wave function
overlap matrix, respectively,  in a subspace consisting of spin-1 spin-wave
excited states\cite{Li2010,DallaPiazza2015}. The projected Hamiltonian system
is diagonalized through the general eigen equation which gives eigenvalues as
the spinon-antispinon excitation energies and spectrum representation for
$S(\mathbf{q},\omega)$\cite{Li2010,DallaPiazza2015}.

The best variational GPWF for the ground state of KAFHM is the Dirac spin
liquid (DSL)\cite{Hastings2000,Ran2007,Iqbal2013}. DSL has flux $\pi$ in the
hexagons of kagome lattice in the mean field Hamiltonian. For comparison, we
also study a zero-flux state which is a spinon Fermi-surface spin liquid (FSL).
If the spin system  in Herbertsmithite doesn't reach a true ground state, a
random-flux spin liquid (RSL) is also possible. We find that all three spin
liquid states have a SWC spectrum in $S(\mathbf{q},\omega)$ with a low
intensity in the elementary BZ and high intensity in 2nd BZ. The spectrum width
of SWC is around $\sim3J$  and the integrated intensity of
$S(\mathbf{q},\omega)$ up to $0.2J$ corresponds to around 20\% of the
equal-time spin structure factor $S(\mathbf{q})$. Unlike one-dimensional (1D)
antiferromagnetic spin-1/2 chain\cite{Mueller1981}, the bottom boundary edge of
SWC is weakly dispersive and the intensity at the edge SWC is not divergent.
Above the low boundary edge, $S(\mathbf{q},\omega)$ is almost energy
independent and weakly depends on the momentum over a wide range of momentum.
These general features of SWC in spin liquids on the kagome lattice are
consistent with experimental observations. FSL has a low energy gap in
$S(\mathbf{q},\omega)$ at the $M$ point and low energy intensity peaks at $K$
points in the magnetic BZ (MBZ).  RSL has a similar $S(\mathbf{q},\omega)$ to
FSL, but the gap at $M$ points is smeared out.

\begin{figure}[b] \includegraphics[width=0.48\columnwidth]{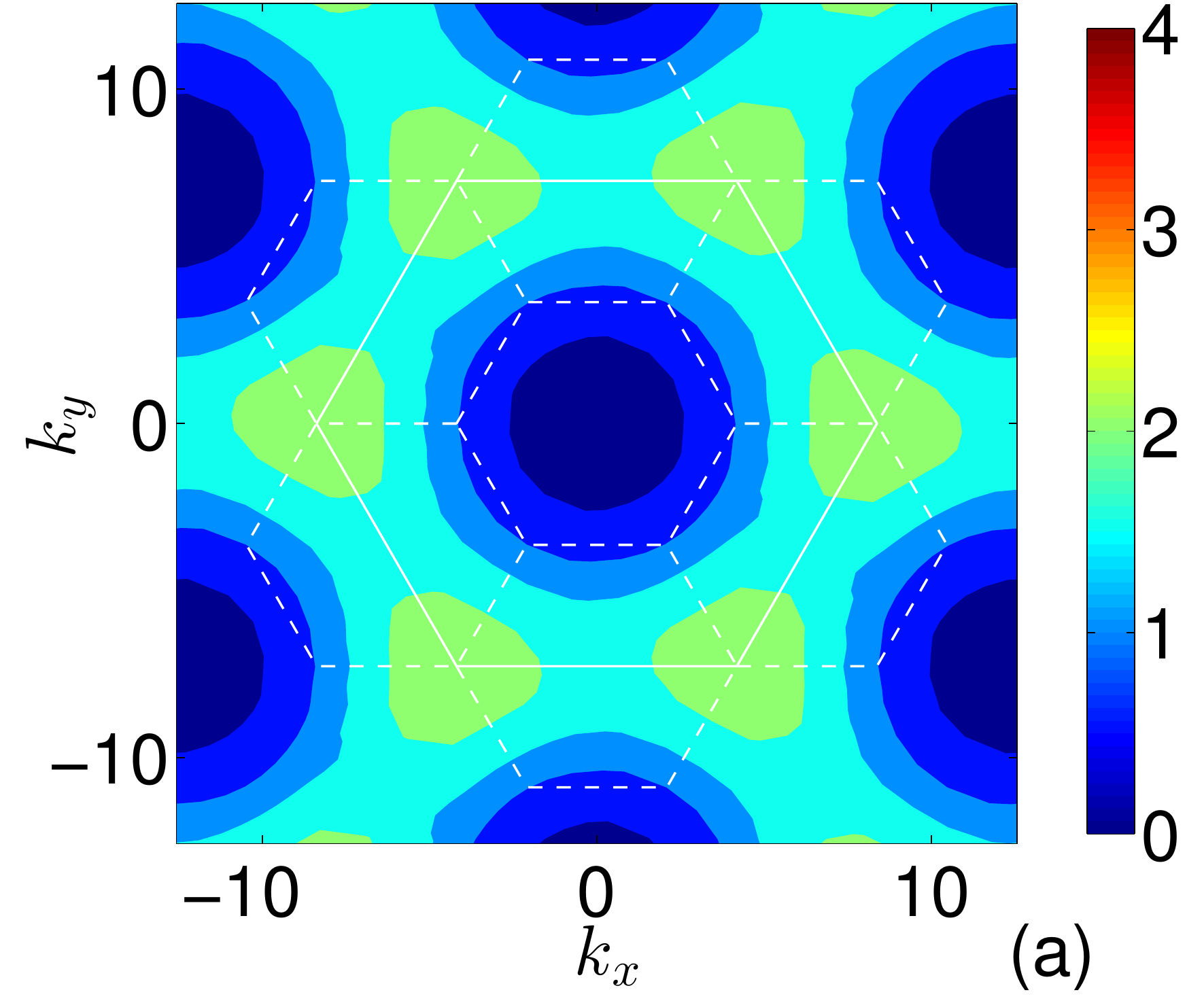}
\includegraphics[width=0.48\columnwidth]{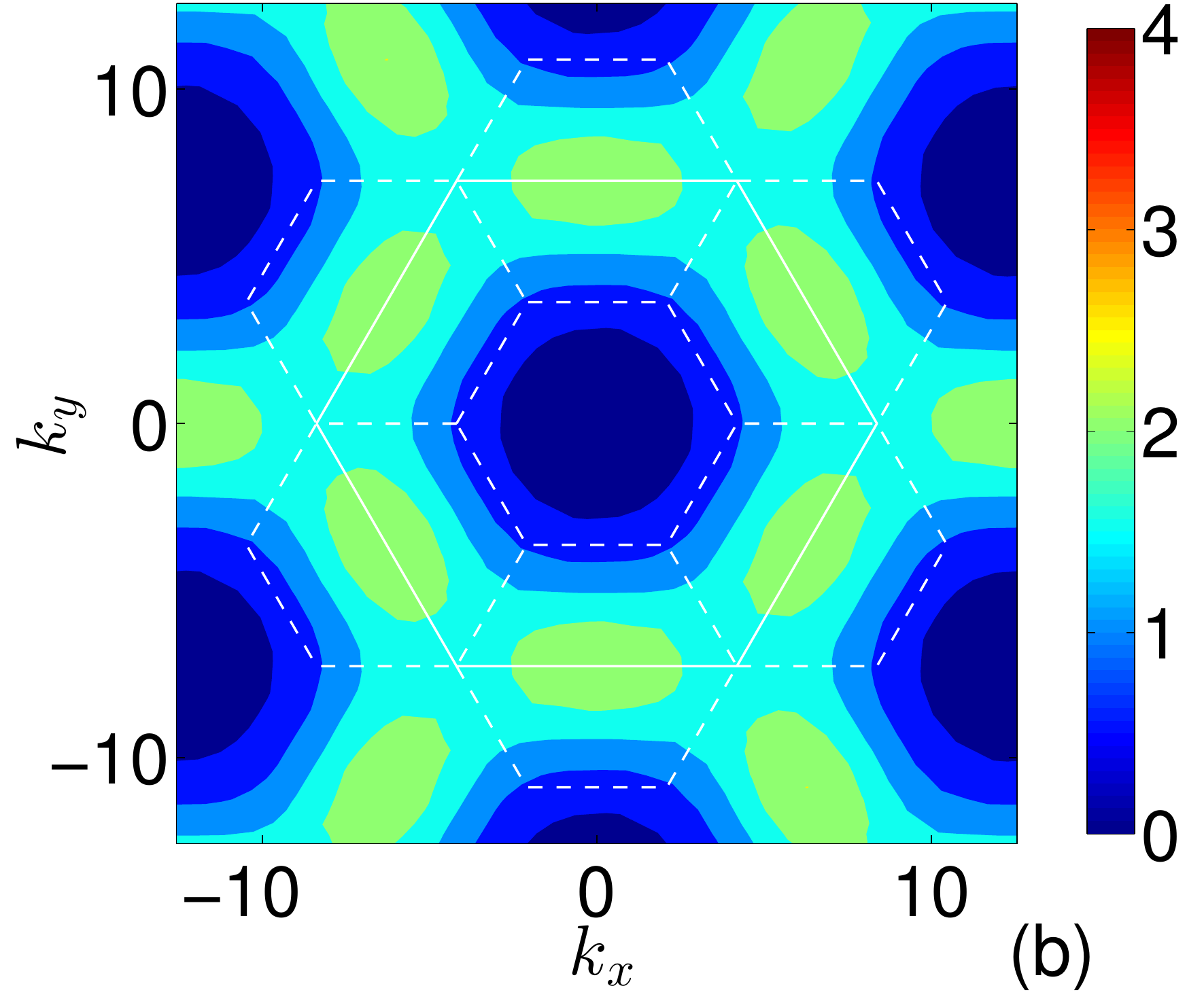}
\includegraphics[width=0.48\columnwidth]{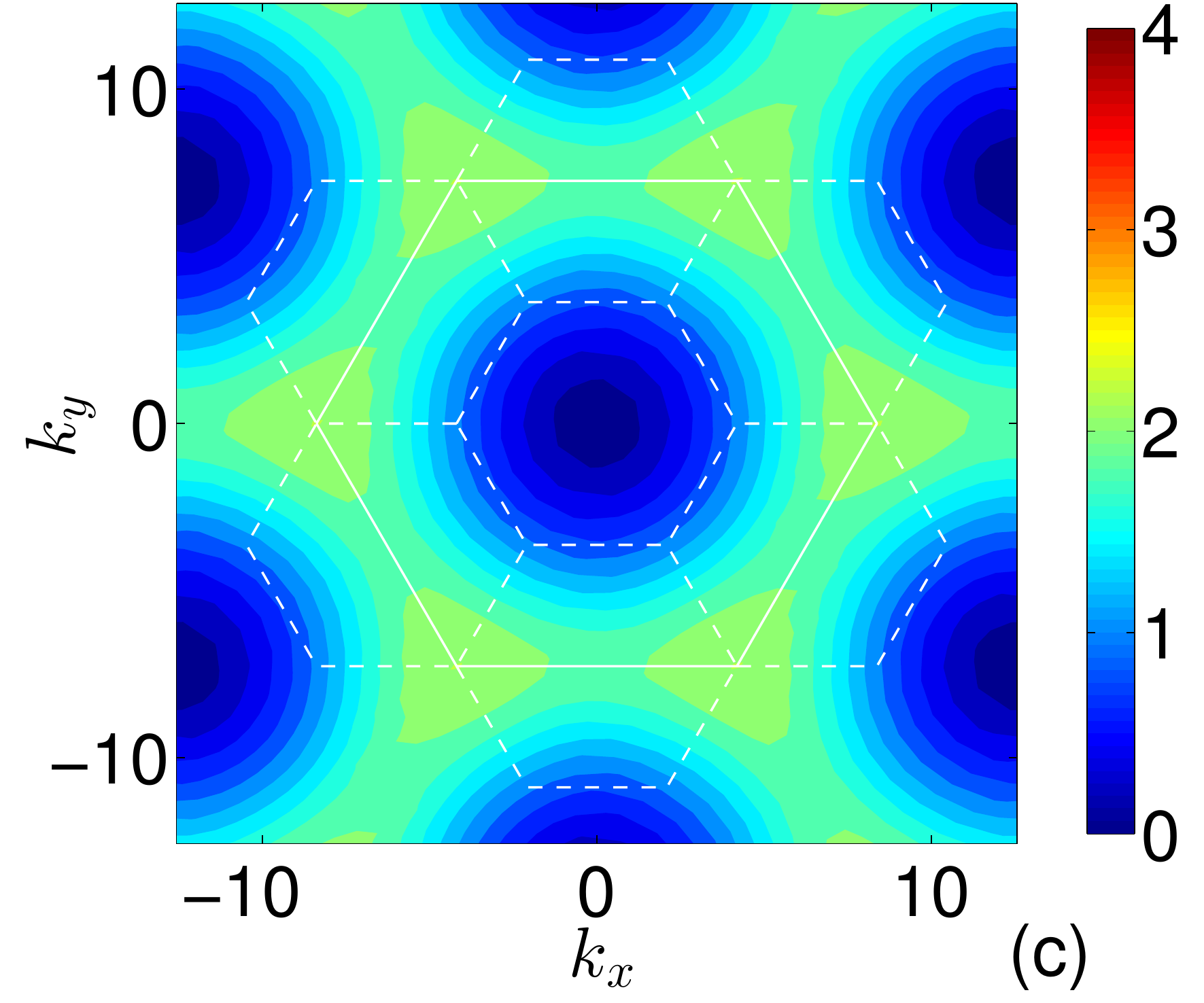}\includegraphics[width=0.5\columnwidth]{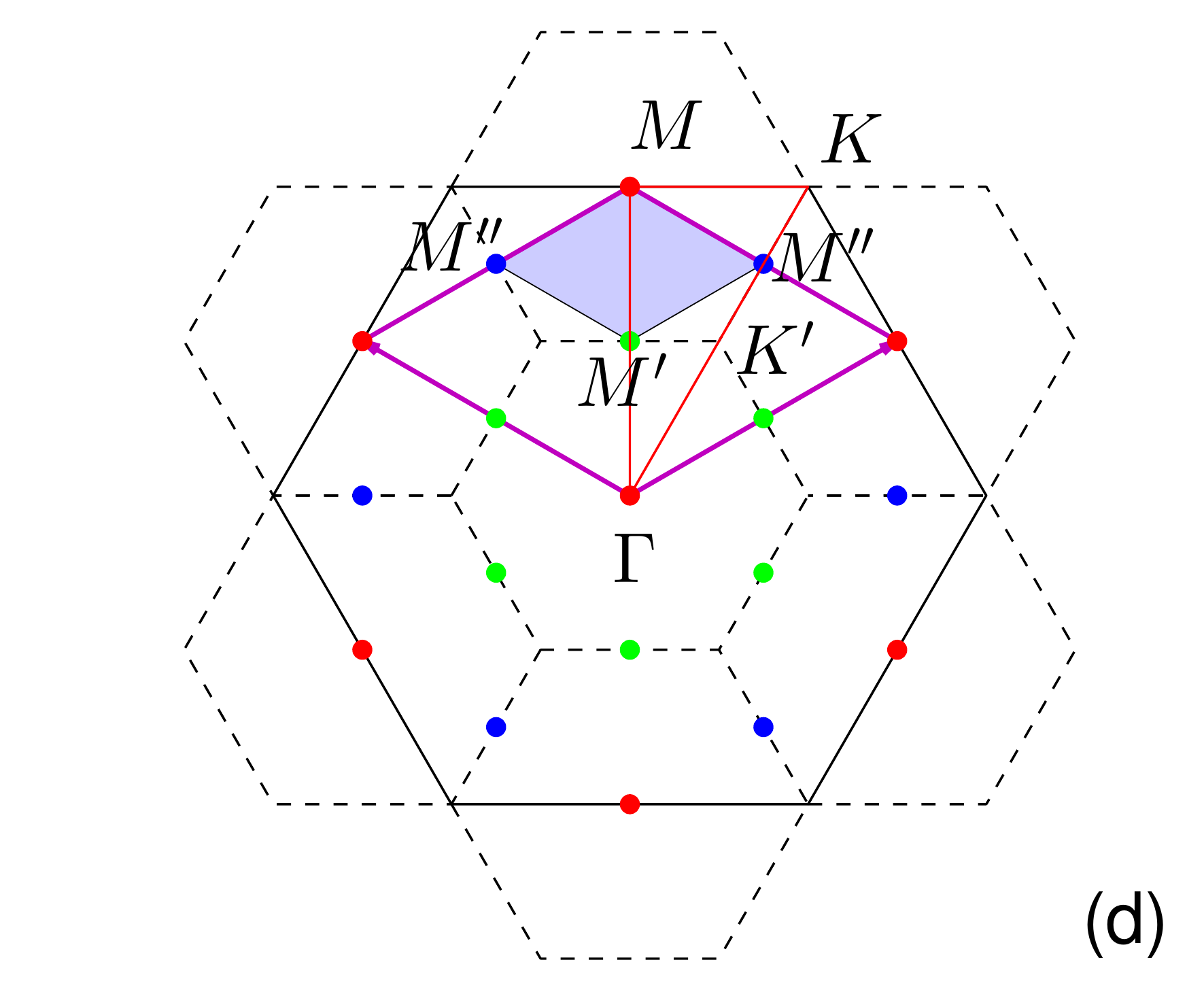}
\includegraphics[width=\columnwidth]{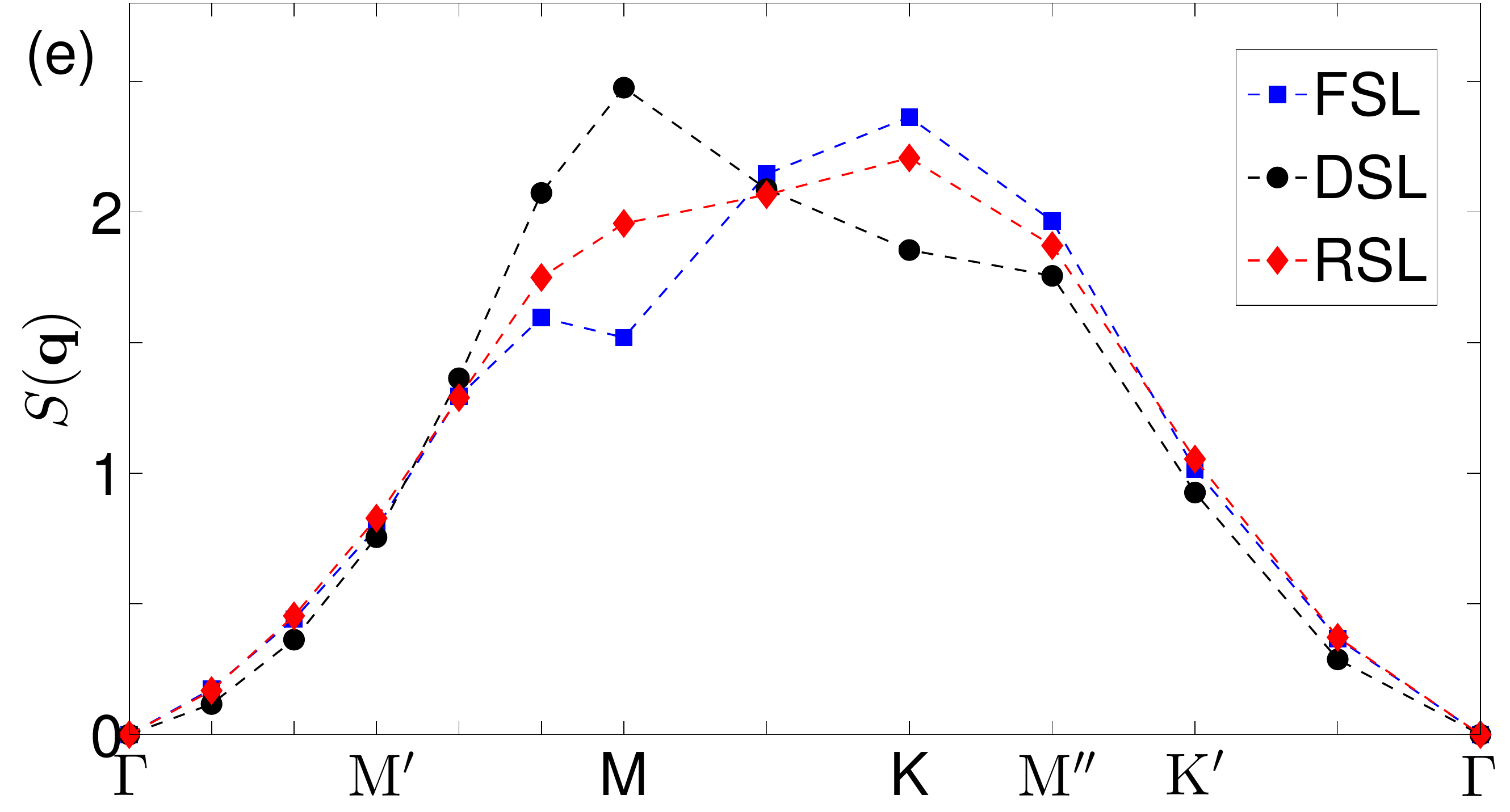} \caption{Contour plots of
equal-time spin structure factor $S(\mathbf{q})$ as a function of momentum in
FSL (a), DSL (b) and RSL (c). The solid and dashed hexagons are the magnetic
extended and elementary Brillouin zones, respectively as shown in (d). Two
spinon dispersion is periodic in the shadow parallelogram in (d). (e)
$S(\mathbf{q})$ along high symmetry directions for FSL, DSL and RSL. }
\label{fig:SF} \end{figure}

In comparison, the low energy intensity peaks of $S(\mathbf{q},\omega)$ in the
experiments\cite{Han2012} are located at $M$ points in the MBZ and high
intensity region connecting $M$ points goes through $M''$ points instead of $K$
points. Therefore, FSL and RSL are not likely to be spin liquid states realized
in Herbertsmithite. DSL has no obvious conflict in $S(\mathbf{q},\omega)$ with
neutron scattering measurements. Particularly, the momentum resolved density of
states for spin-1 excitations in DSL has two cones
at low energies at $M$ and $M''$ points. Different from FSL, spinons in DSL
carry a fractionalized crystal momentum. As a result, the momentum resolved
density of states for spin-1 excitations
is periodic in one-quarter (shadow region in Fig.  \ref{fig:SF} (d)) of the
elementary BZ. The boundary edge below SWC for DSL resembles the mean field
continuum edge and Dirac cones around $M''$ points are due to a crystal
momentum fractionalization. We suggest neutron scattering measurements to
detect low-energy intensity peaks around $M''$ points to experimentally
discover the phenomenon of crystal momentum fractionalization.  This will be a
smoking gun for the DSL in Herbertsmithite.

We start with KAFHM for spins in Herbertsmithite \begin{eqnarray}
\label{eq:KAFH} H=J\sum_{ \langle ij\rangle}\mathbf{S}_i\cdot\mathbf{S}_j,
\end{eqnarray} where $J\sim17$ meV and the summation runs over nearest neighbor
bonds. We will use the Schwinger fermion representation for spin-1/2 operator,
$S_i^a=\frac{1}{2}\sum_{\alpha\beta}f_{i}^\dag\sigma^a f_{i}$. Here
$\sigma^{a=x,y,z}$ are Pauli matrices. The fermionic spinon operator
$f_{i\sigma}$ describes a spin physical Hilbert space within
one-particle-per-site constraint $\sum_{\alpha}f_{i\alpha}^\dag f_{i\alpha} =
1$.

A spin liquid state is characterized by a mean-field Hamiltonian
\begin{eqnarray} \label{eq:MF} H_{\text{MF}}=-\sum_{\langle
ij\rangle}(\chi_{ij}f_{i\sigma}^\dag f_{j\sigma}+\text{H.C.}).  \end{eqnarray}
The GPWFs for a spin liquid ground state and spin-1 excited states are written
as \begin{eqnarray} \label{eq:GPWF}
|\Psi\rangle=\mathcal{P}_G|\Psi_{\text{MF}}^{\chi_{ij}}\rangle,\quad
|\Psi^{S=1}_{ij}\rangle=\mathcal{P}_Gf_{e_i\uparrow}^\dag
f_{e_j\downarrow}|\Psi_{\text{MF}}^{\chi_{ij}}\rangle, \end{eqnarray} where
$\mathcal{P}_G$ is the Gutzwiller projection operator to enforce
one-particle-per-site constraint and $|\Psi_{\text{MF}}^{\chi_{ij}}\rangle$ is
the mean field ground state. $f_{e_i\sigma}$ is the operator for the wave
packet with mean field energy level $e_i$ in the mean field Hamiltonian.

Different choices of $\chi_{ij}$ in Eq. (\ref{eq:MF}) give us different spin
liquid states. FSL is a zero-flux state and has a large spinon Fermi surface.
DSL has flux $\pi$ in the hexagons of kagome lattice. RSL has a random quenched
gauge field $a_{ij}$ on the bond, $\chi_{ij}=|\chi_{ij}|e^{ia_{ij}}$ with
$-\pi\leq a_{ij}\leq \pi$ randomly.

From the ground-state GPWF, we can calculate the equal-time spin structure
factor $S(\mathbf{q})$ \begin{eqnarray} \label{eq:ETSF}
S(\mathbf{q})=\frac{1}{N}\sum_{ij}e^{i\mathbf{q}\cdot\mathbf{r}_{ij}} \langle
S_i^- S_j^+\rangle_0, \end{eqnarray} where
$\mathbf{r}_{ij}=\mathbf{r}_i-\mathbf{r}_j$ and the position summation runs
over all sites on the kagome lattice. $\langle\cdots\rangle_0$ is average over
spin configurations in the ground-state GPWF. For a $12\times6\times3$ lattice,
we specify the general periodic boundary conditions on the lattice
\begin{eqnarray} \label{eq:PBC} f_{i+L_x}=f_{i},\quad f_{i+L_y}=f_{i}e^{ik_0},
\quad k_0=\sqrt{\pi}/2, \end{eqnarray} to get a full shell of mean field energy
levels.  The kagome lattice has the primitive basis
$\mathbf{a}_{1,2}=\pm\frac{1}{2}\hat{\mathbf{e}}_x+\frac{\sqrt{3}}{2}\hat{\mathbf{e}}_y$.
The reciprocal primitive vectors are
$\mathbf{g}_{1,2}=\pm2\pi\hat{\mathbf{k}}_{x}+\frac{2\pi}{\sqrt{3}}\hat{\mathbf{k}}_y$
indicated by the purple parallelogram in Fig. \ref{fig:SF} (d).
$S(\mathbf{q})$ is periodic in extended MBZ (solid hexagons in
Fig.\ref{fig:SF}).

In Fig. \ref{fig:SF} (a), (b) and (c), we compare $S(\mathbf{q})$ among FSL,
DSL and RSL. $S(\mathbf{q})$ has a similar  overall feature for three spin
liquid states. The main differences are the peak positions: While FSL and RSL
have peaks around $K$ points, DSL has peaks at $M$ points in MBZ. As shown in
Fig. \ref{fig:SF} (e), along high symmetry directions in MBZ, FSL has a dip
around $M$ points and the dip feature is smeared out in RSL. DSL has a kink
around $M''$ points. RSL has no translational symmetry and $S(\mathbf{q})$ is
obtained as $\mathbf{q}$-Fourier transformation in Eq. (\ref{eq:ETSF}) on a
$12\times6\times3$ lattice.

\begin{figure}[b] \centering \includegraphics[width=\columnwidth]{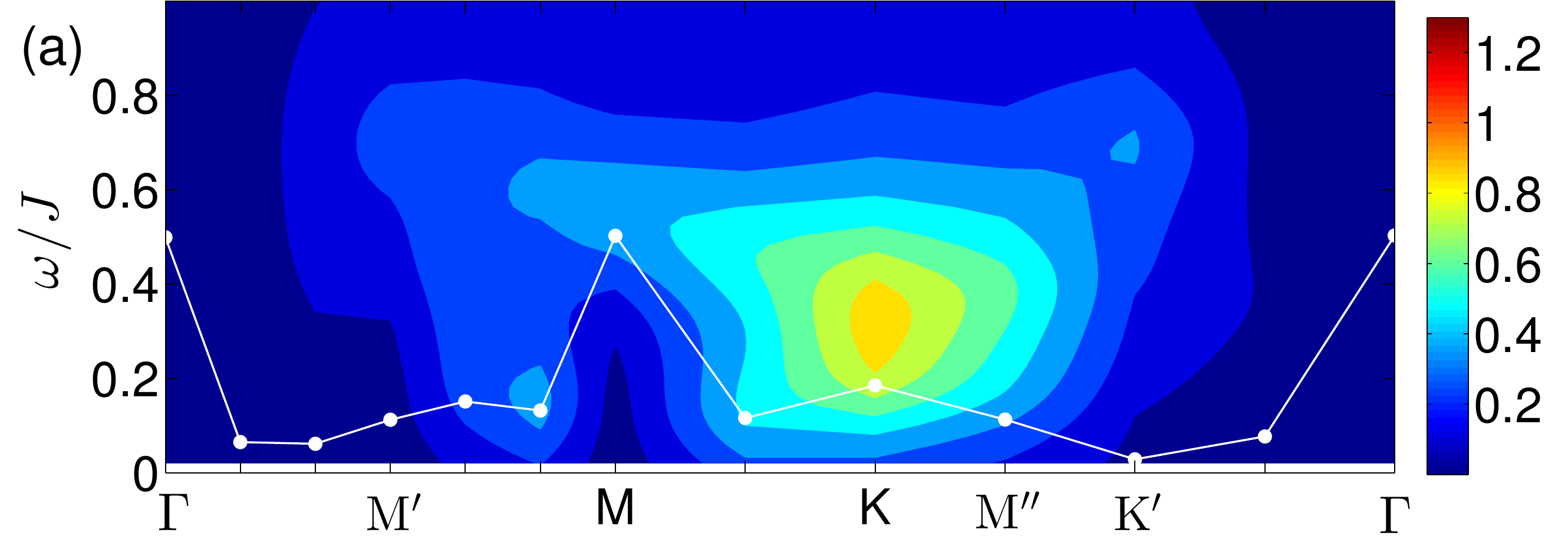}
\includegraphics[width=\columnwidth]{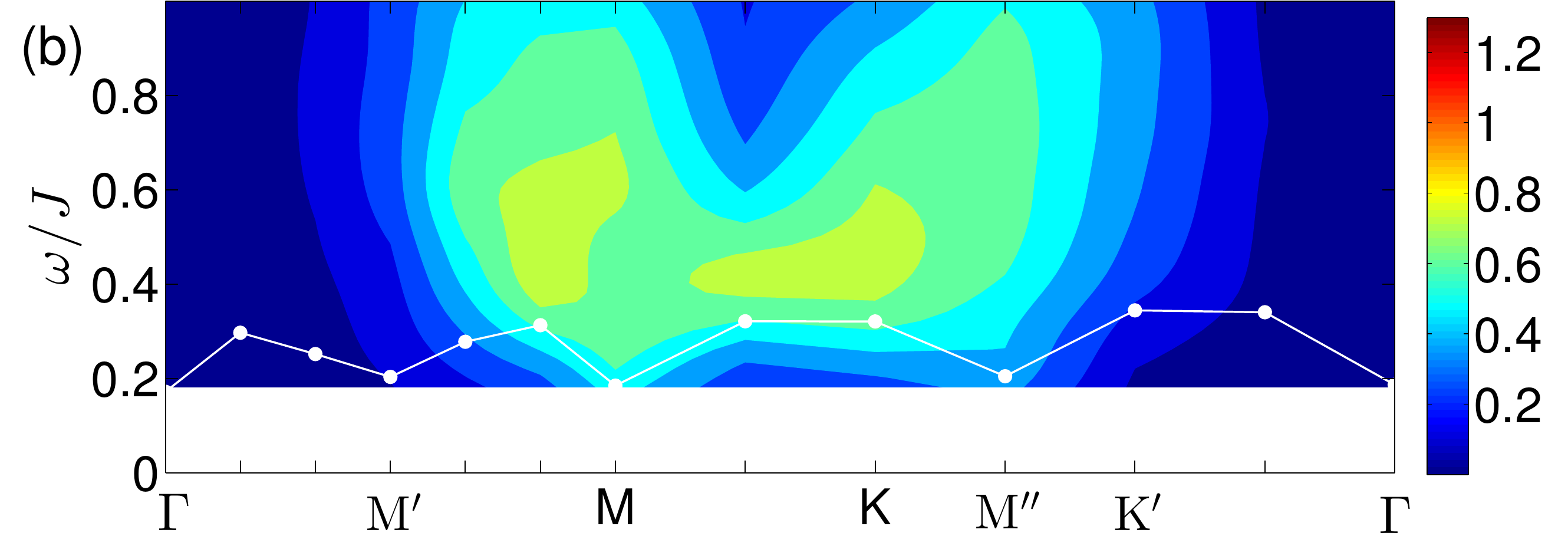}
\includegraphics[width=0.48\columnwidth]{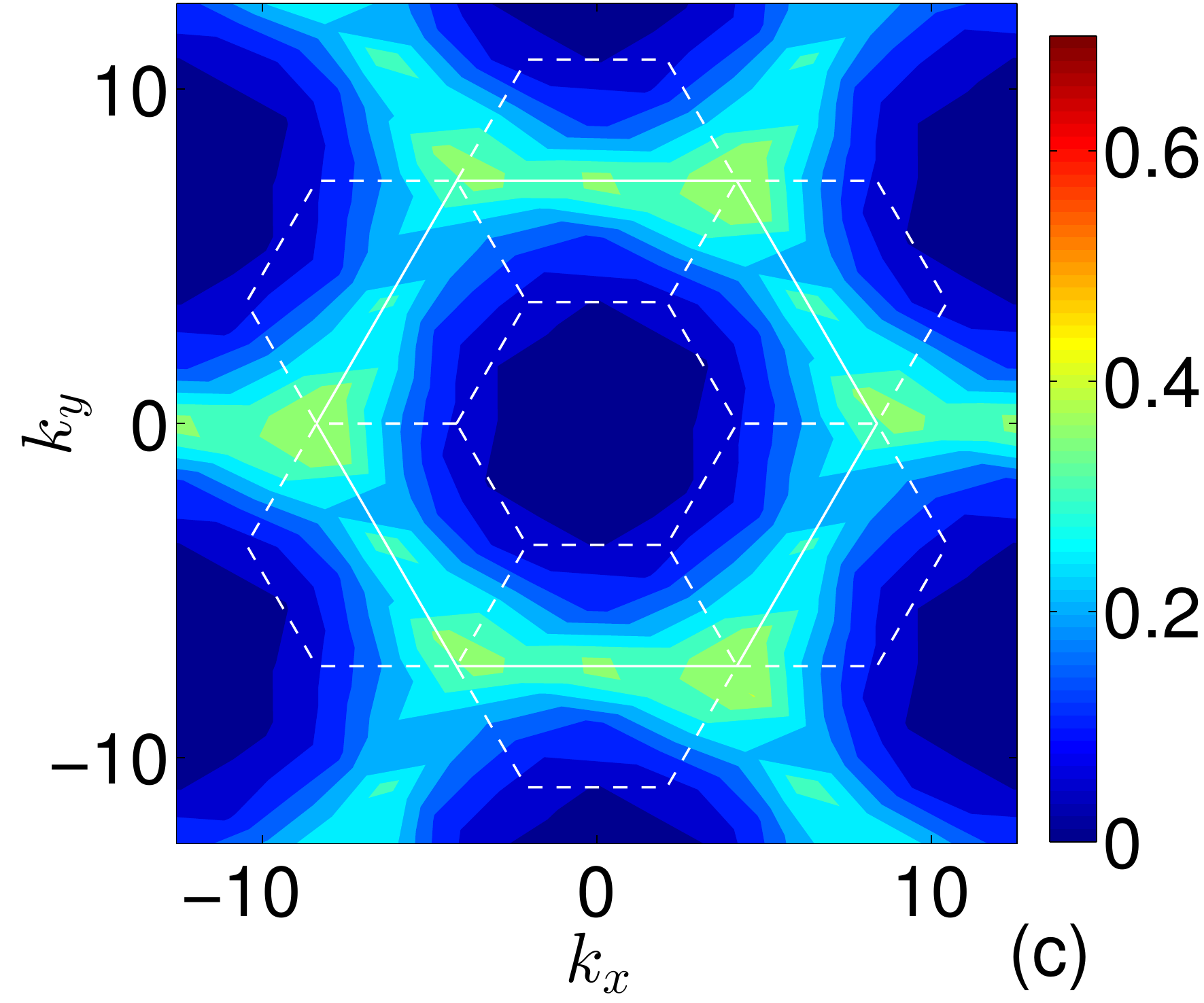}
\includegraphics[width=0.48\columnwidth]{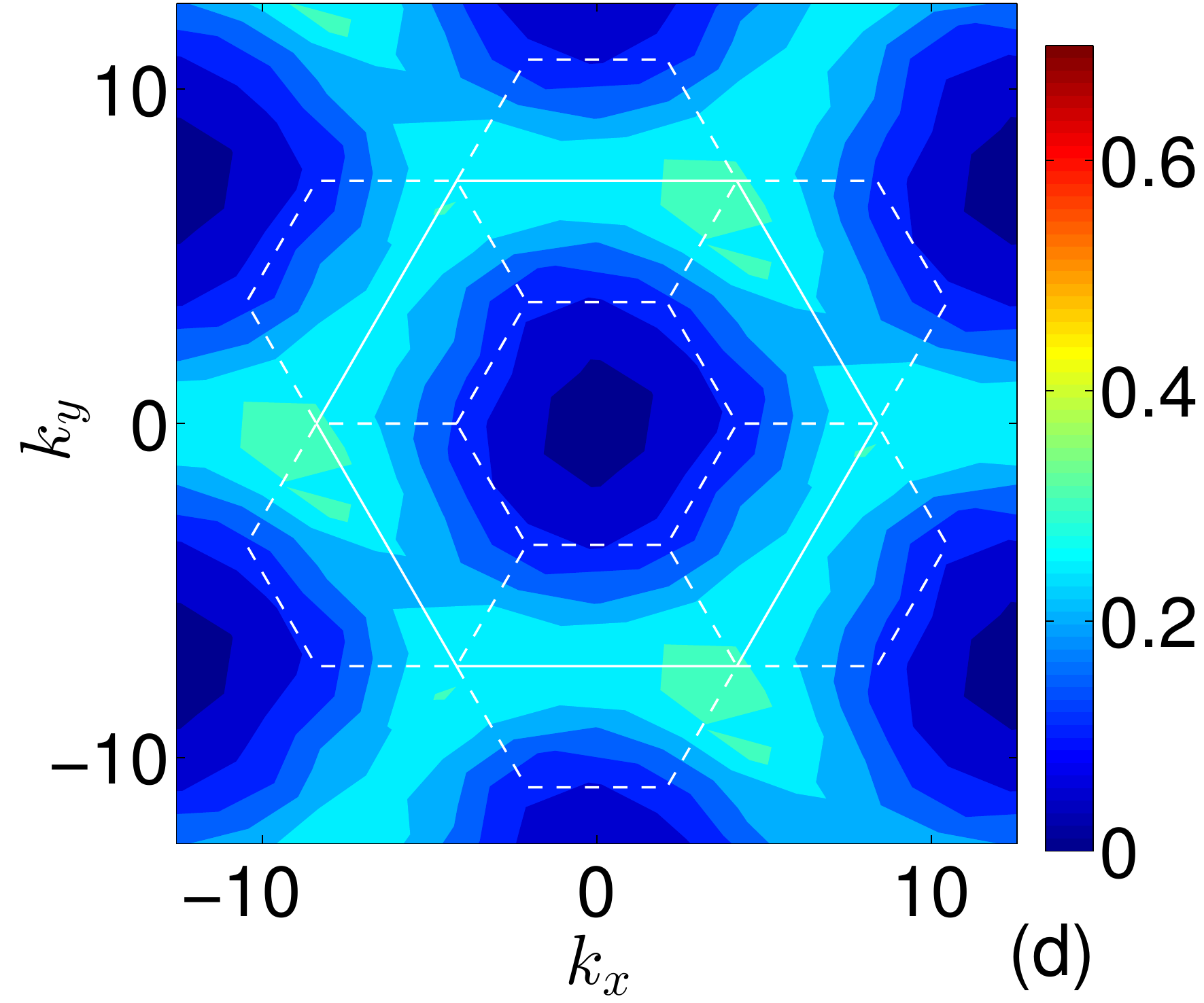} \caption{(a) and (b)
are contour plots of $S(\mathbf{q},\omega)$ in FSL (a) and DSL (b) as a
function of frequency and momentum  along high symmetry directions. The white
solid circles are the lower edge $E_\text{edge}(\mathbf{q})$ of the SWC. (c)
and (d) are contour plots of $S(\mathbf{q},\omega)$ for RSL with fixed energies
$\omega=0.05J$ (c) and $\omega = 0.5J$ (d) as a function of momentum.}
\label{fig:SD} \end{figure} \begin{figure}[t] \centering
\includegraphics[width=0.9\columnwidth]{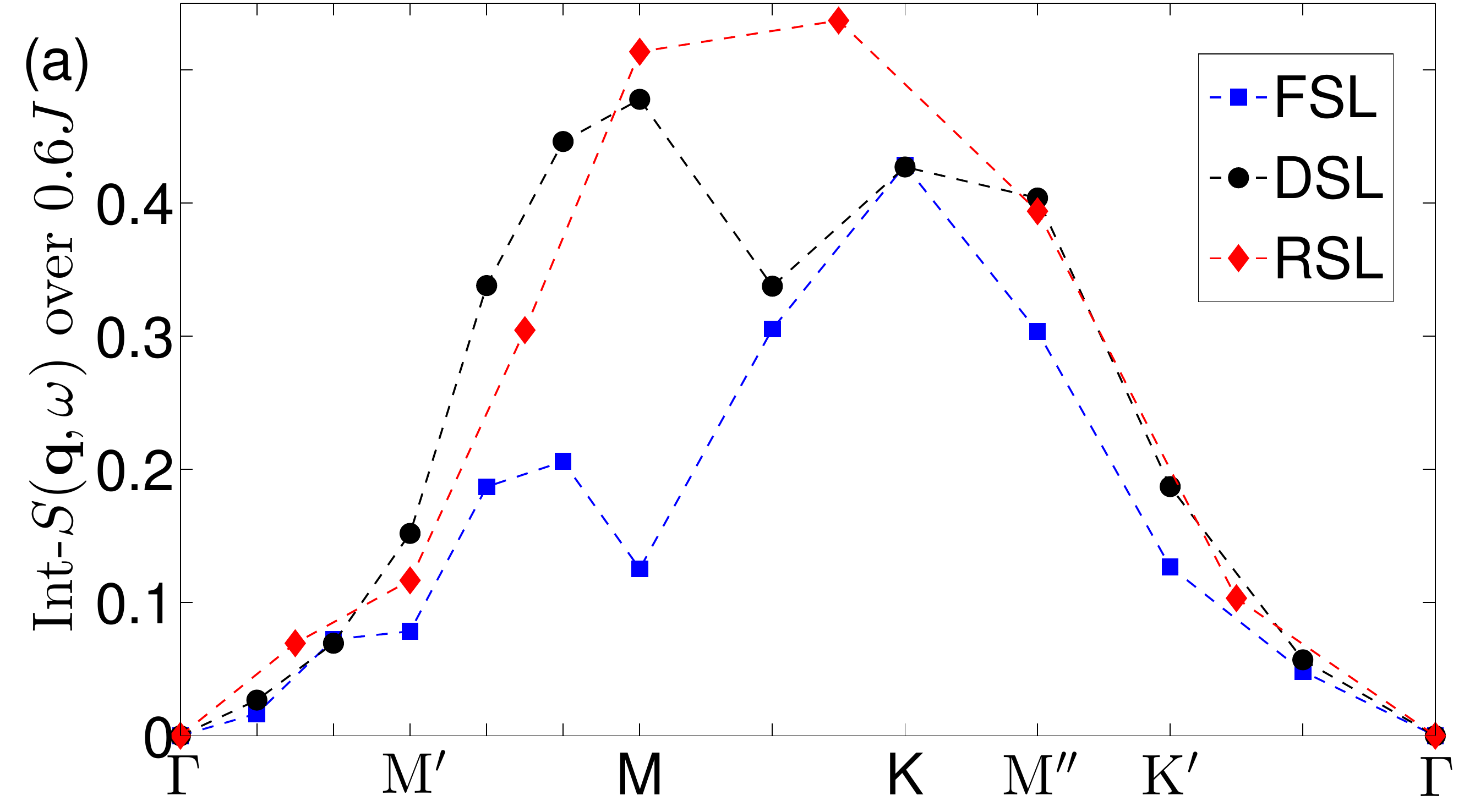}
\includegraphics[width=0.9\columnwidth]{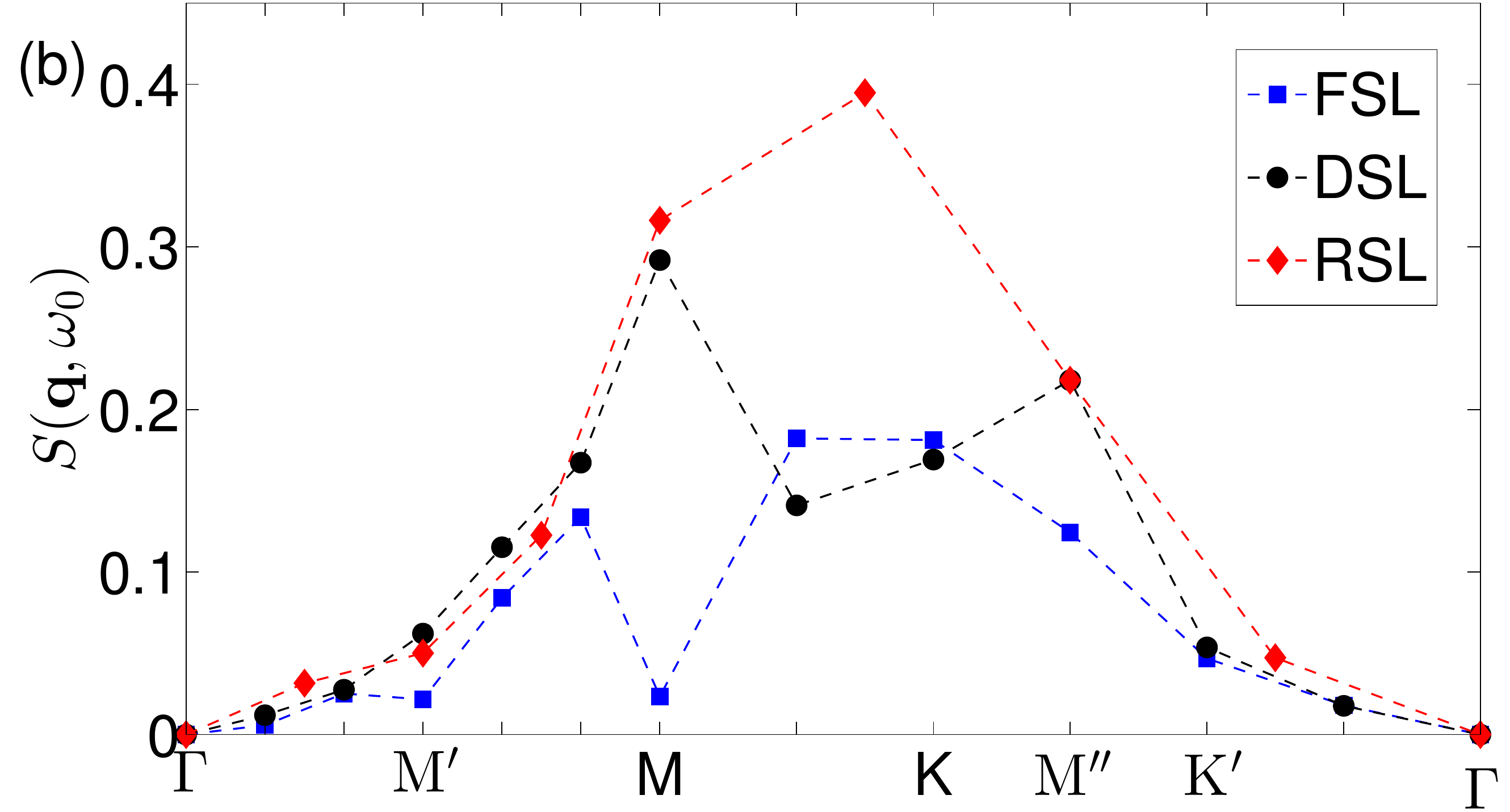} \caption{ (a) Plot of
integrated $S(\mathbf{q},\omega)$ up to $0.6J$ and (b) plot of
$S(\mathbf{q},\omega)$ with fixed frequency $\omega_0 = 0.03J,0.18J,0.01J$ for
FSL, DSL and RSL, as a function of momentum along high symmetry directions. }
\label{fig:SDMTKS} \end{figure}

Neutron scattering experiments measure the dynamic structure factor
\begin{eqnarray} \label{eq:spincorr} S(\mathbf{q},i\omega_n)=\int_0^\beta d\tau
e^{i\omega_n\tau}\frac{1}{N}\sum_{ij}e^{i\mathbf{q}\cdot\mathbf{r}_{ij}}
\langle T_\tau S_i^-(\tau)S_j^+(0)\rangle_0.\nonumber \end{eqnarray} The
projected Hamiltonian system within a subspace consisting of spin-1 excited
states\cite{Li2010,Note1} is given as \begin{eqnarray} \label{eq:projH}
\mathbb{H}(i'j',ij) = \langle i'j'|H|ij\rangle,\quad\mathbb{O}(i'j',ij) =
\langle i'j'|ij\rangle, \end{eqnarray} where $|ij\rangle$ is
$|\Psi^{S=1}_{ij}\rangle$ in Eq. (\ref{eq:GPWF}). The matrix elements in Eq.
(\ref{eq:projH}) are evaluated by using Monte Carlo methods\cite{Li2010,Note1}.
The projected Hamiltonian system $\{\mathbb{H},\mathbb{O}\}$ is diagonalized
through a generalized eigen equation, \begin{eqnarray} \label{eq:GEE}
\mathbb{H}|\phi_n\rangle = \epsilon_n\mathbb{O}|\phi_n\rangle, \end{eqnarray}
where $|\phi_n\rangle$ and $\epsilon_n$ are spin-1 two-spinon wave functions
and energy levels, respectively. In terms of them, we has the spectral
representation \begin{eqnarray} \label{eq:SDF}
S(\mathbf{q},\omega)=\sum_n\delta(\omega-(\epsilon_n-\epsilon_0))|\langle\phi_n|S_{\mathbf{q}}^+\mathcal{P}_G|\Psi^{\chi_{ij}}_{\text{MF}}\rangle|^2,
\end{eqnarray} where $\epsilon_0$ is the ground state variational
energy\footnote{See supplementary materials for details.}.

FSL and DSL have translational symmetry and the projected Hamiltonian system
$\{\mathbb{H},\mathbb{O}\}$ will be labeled  according to the momentum in a
relatively large system ($12\times6\times3$).  The mean field dispersion for
spinons has a finite-size gap on  the $12\times6\times3$ lattice with the
boundary conditions in Eq. (\ref{eq:PBC}). The finite-size spin gaps are
$E_{\text{MF}}^{\text{sg}}=0.029|\chi|$  and
$E_{\text{MF}}^{\text{sg}}=0.586|\chi|$ for FSL and DSL, respectively. Here
$|\chi|$ is the mean field spinon hopping amplitude. Correspondingly, spin-1
excitations have a gap, $E_g=0.03J$ and $E_g=0.18J$ for FSL and DSL,
respectively.   Contour plots of $S(\mathbf{q},\omega)$ are shown in Fig.
\ref{fig:SD} (a) FSL and (b) DSL, with broadening $\eta = 0.15J$ for delta
function in Eq. (\ref{eq:SDF}),
$\delta(\omega-\epsilon)\rightarrow\frac{\eta/\pi}{(\omega-\epsilon)^2+\eta^2}$.
The RSL state has no translational symmetry and the computation complexity
increases considerably. $S(\mathbf{q},\omega)$ for RSL is computed in the whole
Brillouin zone only on a $4\times4\times3$ lattice. In Fig.\ref{fig:SD}, we
plot $S(\mathbf{q},\omega)$ for RSL with fixed frequencies (c) $\omega=0.05J$
and (d) $\omega = 0.5J$.

The projected Hamiltonian system $\{\mathbb{H},\mathbb{O}\}$ for spin-1 excited
states has the SWC width around $W_\text{swc}\simeq3J$ for three spin  liquid
states. In Fig. \ref{fig:SDMTKS} (a), we plot the integrated
$S(\mathbf{q},\omega)$ over 0 to $0.6J$ along high symmetry directions.  Note
that Fig. \ref{fig:SF} (e) and Fig. \ref{fig:SDMTKS} (a) has the same unit. The
integrated $S(\mathbf{q},\omega)$ over 0 to $0.6J$ has about 20\% intensity of
the fully integrated of spin spectral weight $S(\mathbf{q})$. To explore the
low energy features, we plot $S(\mathbf{q},\omega)$ with fixed energies a
little higher above the finite-size gap, $\omega_0=0.03J,0.18J,0.01J$ for FSL,
DSL and RSL,  along high symmetry directions. The low energy cones at $M$ and
$M''$ points are very unique for DSL and clearly resolved in Fig.
\ref{fig:SDMTKS} (b).

\begin{figure}[t] \centering
\includegraphics[width=0.9\columnwidth]{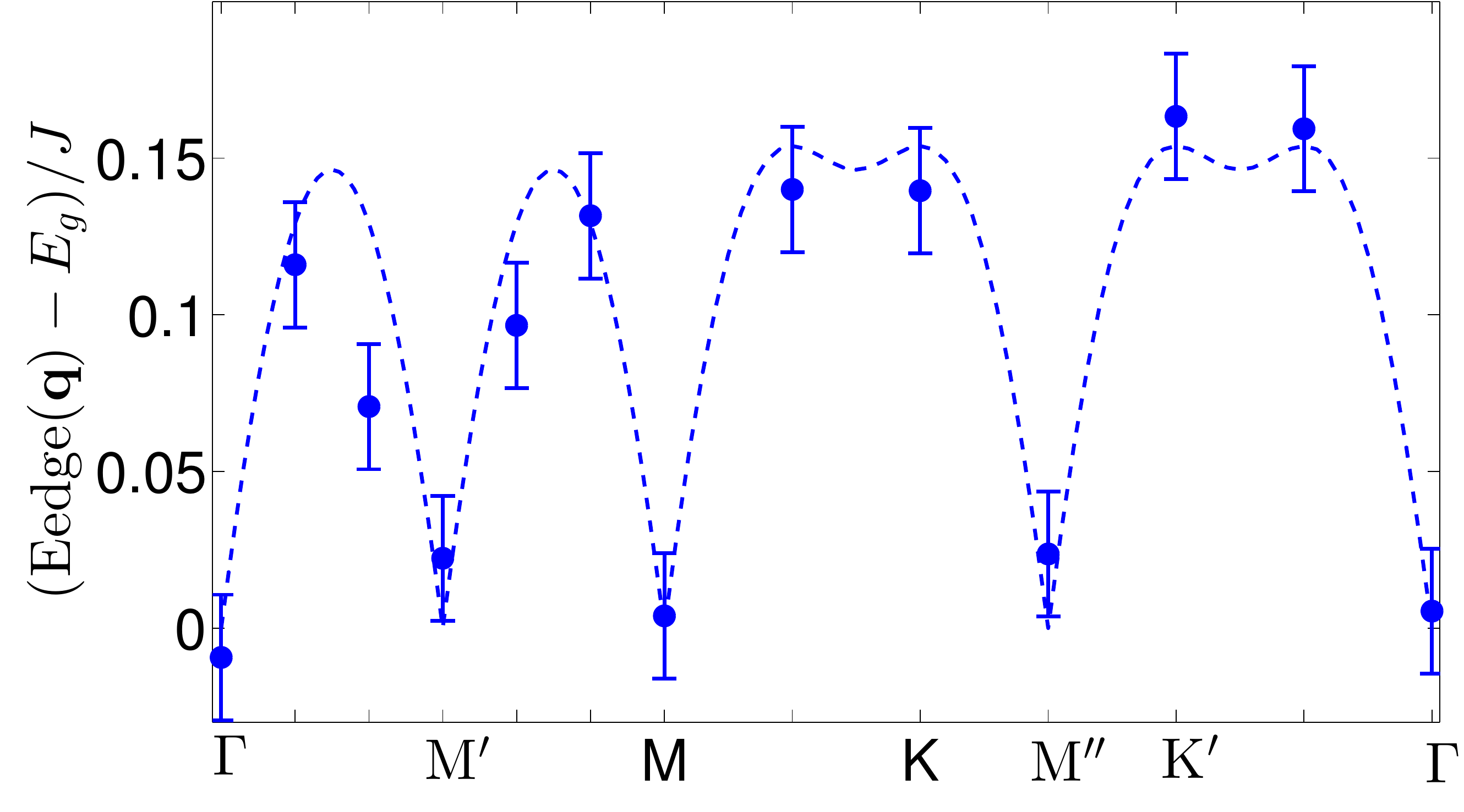} \caption{DSL:  the lower
edge of SWC, $E_\text{edge}(\mathbf{q})-E_g$, as a function of momentum along
high symmetry directions. $E_g$ is the finite-size gap. The dashed line is the
lower edge of SWC from the meanfield theory with $|\chi|=0.43J$ for hoping
amplitude.} \label{fig:edge} \end{figure}

We can decompose $S(\mathbf{q},\omega)$ into a spin matrix element
$M(\mathbf{q},\omega)$ and a density of two spinon excited states
\begin{eqnarray} \label{eq:spin_matrix_density}
S(\mathbf{q},\omega)=M(\mathbf{q},\omega) D(\mathbf{q},\omega), \end{eqnarray}
with
$D(\mathbf{q},\omega)=\sum_{n}\delta(\omega-(\epsilon_n(\mathbf{q})-\epsilon_0))$,
where $\epsilon_n(\mathbf{q})$ is the generalized eigen value of the projected
Hamiltonian system $\{\mathbb{H},\mathbb{O}\}$ for a given momentum
$\mathbf{q}$. The lowest generalized eigen values $\epsilon_1(\mathbf{q})$
gives the lower edge of the SWC, $E_{\text{edge}}(\mathbf{q}) =
\epsilon_1(\mathbf{q}) - \epsilon_0$, which is plotted as white solid circles
in Fig. \ref{fig:SD} (a) FSL and (b) DSL.

FSL has a large spinon Fermi surface and its low energy sectors of spin-1
excited states strongly depend on the finite size. In the thermodynamic limit
$K$ and $K'$ points are equivalent under 60\degree~rotation symmetry for the
projected Hamiltonian system $\{\mathbb{H},\mathbb{O}\}$. However, in Fig.
\ref{fig:SD} (a), the lower edge at $K$ has a higher energy than $K'$,
$E_\text{edge}(K)>E_\text{edge}(K')$ since the lattice shape
($12\times6\times3)$ and the boundary conditions in Eq. (\ref{eq:PBC}) break
the rotation symmetry. The lower edge for DSL resembles that in the mean field
calculations. In Fig. \ref{fig:edge}, The lower edge of the SWC
$E_\text{edge}(\mathbf{q})-E_g$ for DSL fits well the mean field calculation
with  a finite-size spin gap $E_g=0.18J$ and the mean field hopping amplitude
$|\chi|=0.43J$.

Different from FSL, fermionic spinons in DSL carry a crystal momentum
fractionalization\cite{Wen2002,Lu2011}. Due to $\pi$ flux in the hexagons of
kagome lattice, translational operators for spinons along the primitive lattice
vectors $\mathbf{a}_{1,2}$ anti-commute with each other \begin{eqnarray}
\label{eq:T12} T_1T_2=-T_2T_1,\quad
T_{1,2}(\mathbf{x})=\mathbf{x}+\mathbf{a}_{1,2}.  \end{eqnarray} As a result,
the projected Hamiltonian system $\{\mathbb{H},\mathbb{O}\}$ has the spin-1
spectrum $\epsilon_n$ with a period of one-quarter of the Brillouin zone.
In other words, the momentum
resolved density of states for spin-1 excitation continuum ({\it i.e.}
two-spinon excitations) has
a period of one-quarter of the Brillouin zone.\cite{Wen2002,Wen2004}.
Note that $M'$ and $M''$ points are equivalent to $\Gamma$ points and cut the
elementary BZ (shadow parallelogram in Fig. \ref{fig:SF} (d))  into four
pieces.

The spin matrix element $M(\mathbf{q},\omega)$ in
Eq.(\ref{eq:spin_matrix_density}) is periodic in the MBZ and $M,M',M''$ and
$\Gamma$ are not equivalent any more in $S(\mathbf{q},\omega)$. While the
magnetic intensity at $\Gamma$ and $M'$ points are suppressed in
$S(\mathbf{q},\omega)$ as shown in Fig. \ref{fig:SD}, $M$ and $M''$ are still
visible. The low energy intensity at $M$ and $M''$ in $S(\mathbf{q},\omega)$ is
the implication of a crystal momentum fractionalization in DSL which is
detectable in the neutron scattering measurements.

Here we make several remarks on comparison between experiments and theoretical
results. The three different spin liquids state have a similar overall shape in
$S(\mathbf{q},\omega)$ with general features: a SWC spectrum over large energy
$\sim3J$ with low intensity in the elementary BZ and high intensity in 2nd BZ,
in good agreement with experimental observations. In  one-dimensional
antiferromagnetic spin-1/2 chain,   $D(q,\omega)$ is finite at the lower
boundary and $S(q,\omega)$ has a divergent sharp lower edge due to
$M(q,\omega)$\cite{Mueller1981}. Although enhanced at low energies,
$S(\mathbf{q},\omega)$ does not diverge at the lower edge of the SWC in spin
liquid states on the kagome lattice. So nearly invisible lower edge in
Herbertsmithite experiment may not be a big issue.  For FSL,
$S(\mathbf{q},\omega)$ has a gap at the $M$ point in the extended MBZ, in
contrast to experiments. RSL also has low energy intensity peaks at $K$ points
inconsistent with experiments. In contrast, DSL has no obvious conflict with
experimental observations.  Due to a momentum fractionalization,
$S(\mathbf{q},\omega)$ of DSL has two low energy Dirac cones at $M$ and $M''$
points in the MBZ. In the experiments, below 1.5 meV, high intensity $M$ points
in $S(\mathbf{q},\omega)$ are connected through $M''$ points instead of $K$
point. So low energy intensity peaks at $M''$ points due to a crystal momentum
fractionalization may already be observed in experiments; however, these
features are interpreted as the impurity effects\cite{Han2012}. In the presence
of impurities, the system in Herbertsmithite may have low energy gauge field
fluctuations. We find that high intensity peaks at $M$ and $M''$ are stable
against quenched gauge field fluctuations although the low energy boundary edge
below SWC is smeared out\cite{Note1}. Recently, Barlowite with AF ordering
temperature $T_N=15K$ is studied as another kagome
antiferromagnet\cite{Han2014}. Its non-magnetic (Mg or Zn) doped variety is
proposed to has less imperfections than Herbertsmithite\cite{Han2014,Liu2015}.
The new material is promising to clear the impurity issues.

In conclusion, we study the fractional spin-wave continuum in spin liquid
states on the kagome lattice. We find out that DSL describes the experiments in
Herbertsmithite well. Besides a fractionalized spin moment, fermionic spinons
in DSL carry a fractionalized crystal momentum which is also potentially
detectable in further experiments.

J.W. Mei thanks J. Carrasquilla for useful discussions.  X-G. Wen is supported
by NSF Grant No.  DMR-1005541 and NSFC 11274192.  He is also supported by the
BMO Financial Group and the John Templeton Foundation. Research at Perimeter
Institute is supported by the Government of Canada through Industry Canada and
by the Province of Ontario through the Ministry of Research.

\bibliography{SDkagome}
\end{document}


\title{Fractional spin-wave continuum in spin liquid states on the kagome lattice}

\author{Jia-Wei Mei}
\affiliation{Perimeter Institute for Theoretical Physics, Waterloo, Ontario, N2L 2Y5 Canada}
\author{Xiao-Gang Wen}
\affiliation{Department of Physics, Massachusetts Institute of Technology, Cambridge, Massachusetts 02139, USA}
\affiliation{Perimeter Institute for Theoretical Physics, Waterloo, Ontario, N2L 2Y5 Canada}
\date{\today}
\maketitle

\section{Smeared DSL}
In the Herbertsmithite compound, the system may not reach its true ground state due to the presence of impurities and the gauge fluctuations is quenched. We would like to study smeared DSL (SDSL) in this section.

\begin{figure}[b]
  \centering
\includegraphics[width=0.3\columnwidth]{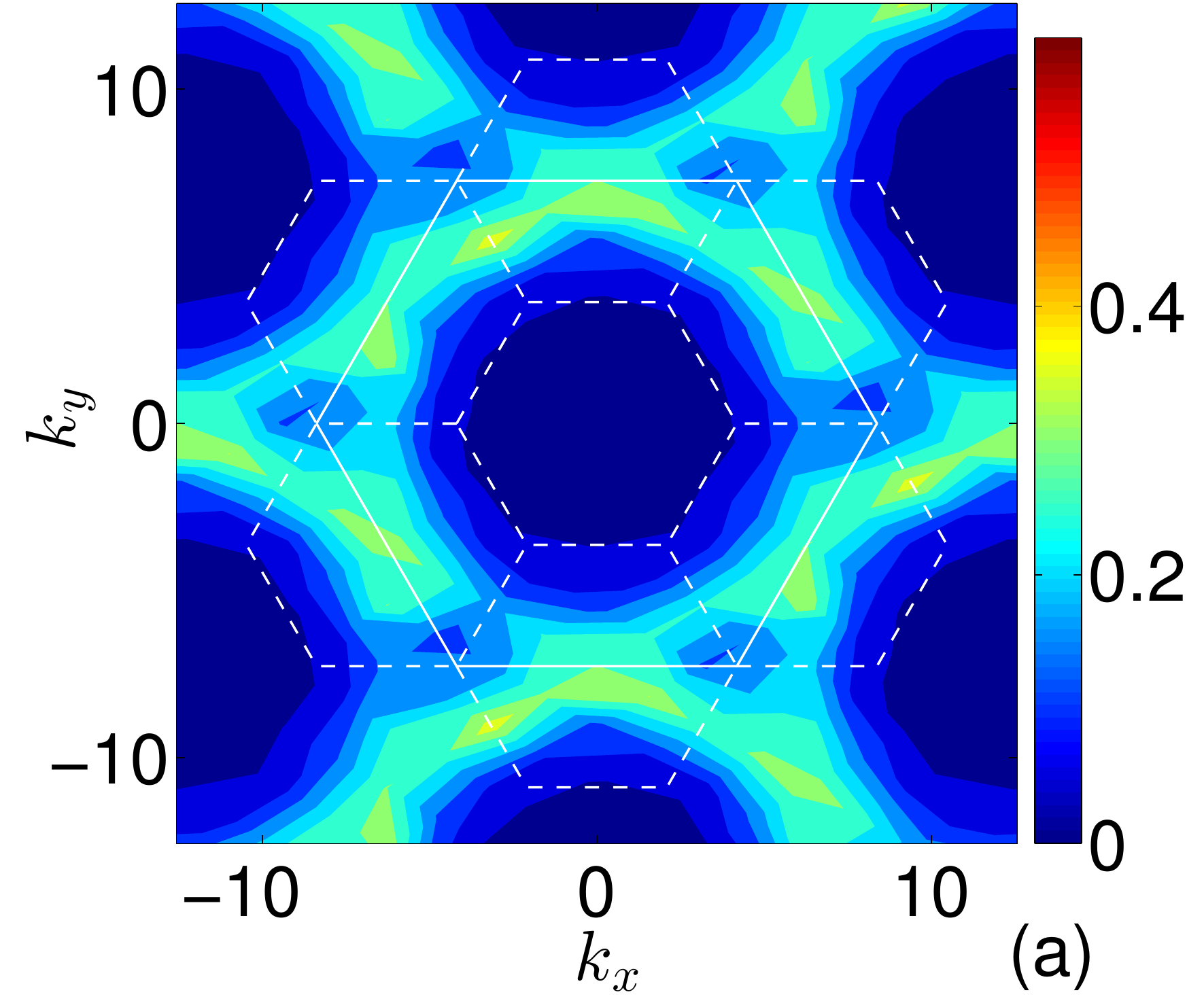} \includegraphics[width=0.3\columnwidth]{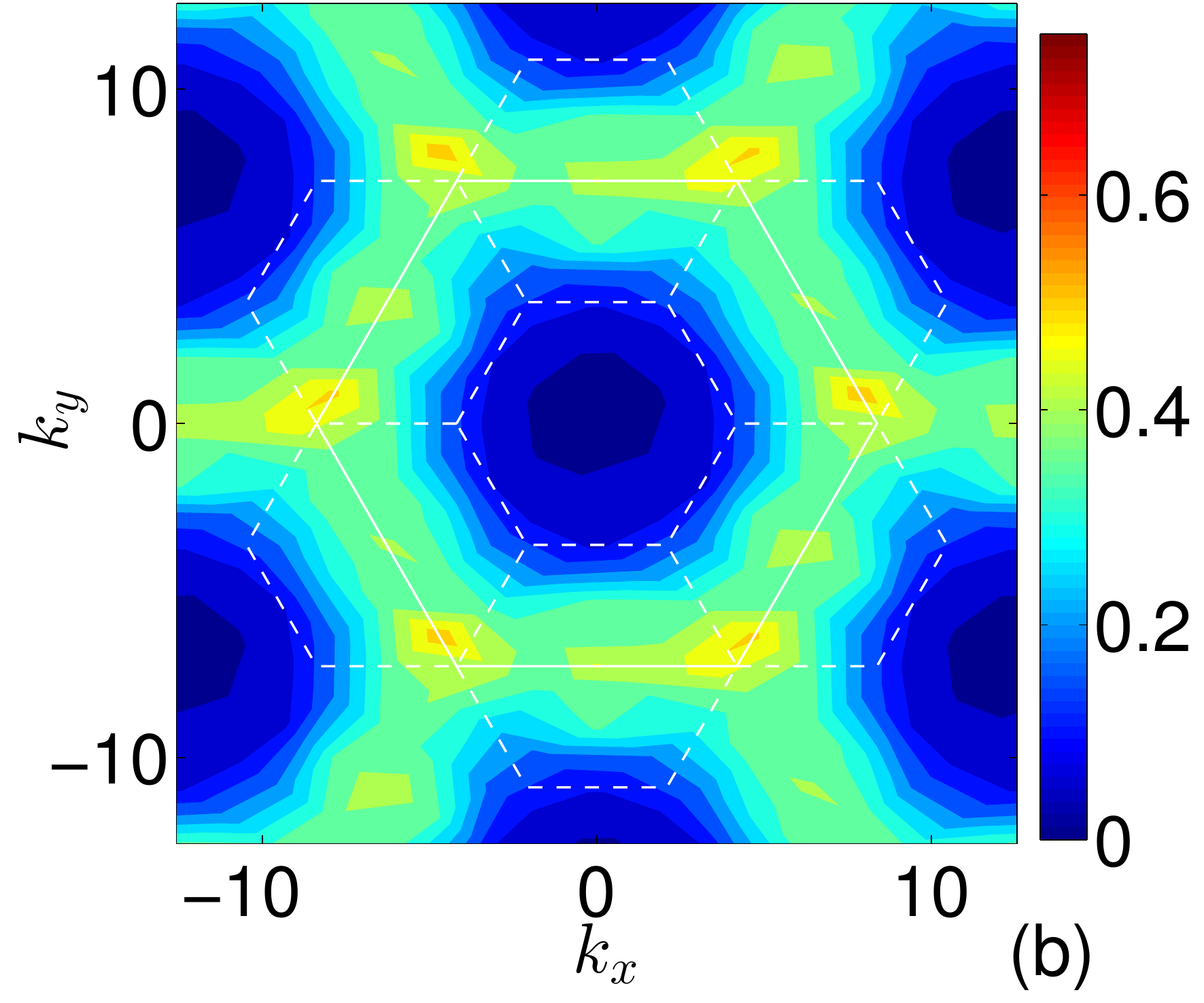}
  \caption{Smeared DSL: Contour plot of dynamic spin structure factor $S(\mathbf{q},\omega)$ as a function of momentum at fixed frequency for (b)$\omega = 0.05J$ and (c) $\omega = 0.5J$. }
  \label{fig:SDSL}
\end{figure}

$\chi_{ij}$ has low-energy gauge fluctuations, $\chi_{ij}=\bar{\chi}_{ij}e^{ia_{ij}}$, where $a_{ij}$ behaves as gauge bosons. The gauge fluctuations are very soft when the ground state of kagome spin model is close to a quantum phase transition. In the Herbertsmithite compound, the system may not reach its true ground state due to the presence of impurities and the gauge fluctuations $a_{ij}$ is quenched. SDSL is described as
\begin{eqnarray}
  \label{eq:GPWFGR}
  |\Psi_{\text{SDSL}}\rangle=\mathcal{P}_G|\Psi_{\text{MF}}^{\bar{\chi}_{ij}a_{ij}}\rangle,
\end{eqnarray}
where $\bar{\chi}_{ij}$ is specified as the DSL and the gauge field $-0.2\pi<a_{ij}<0.2\pi$ is chosen randomly. For the SDSL, GPWFs for excited states are given as
\begin{eqnarray}
  \label{eq:GPWFERSL}
  |\Psi^{S^z=1}(e_i,e_j)\rangle=\mathcal{P}_Gf_{e_i\uparrow}^\dag f_{e_j\downarrow}|\Psi_{\text{MF}}^{\bar{\chi}_{ij}a_{ij}}\rangle.
\end{eqnarray}
The $S(\mathbf{q},\omega)$ is shown in Fig. \ref{fig:SDSL} where the low energy intensity peaks at $M$ and $M''$ points are clear resovled.

\section{Details for MC}
\subsection{GPWFs for Spin liquids}

 The mean field Hamiltonian in Eq. (\ref{eq:MF}) can be easily diagonalized on the finite lattice and it has $N$ energy levels for both spin-up and down spinons, $e_0^{\uparrow/\downarrow}<e_1^{\uparrow/\downarrow}<\cdots<e_{N-1}^{\uparrow/\downarrow}$. Filling the lowest $N/2$ spin-up and $N/2$ spin-down energy levels, we obtain GPWF
\begin{eqnarray}
  \label{eq:gpwf}
  |G\rangle=\mathcal{P}_G\left(e_0^\uparrow\cdots e_{N/2-1}^\uparrow|\text{vac}\rangle\otimes e_0^\downarrow\cdots e_{N/2-1}^\downarrow|\text{vac}\rangle\right),
\end{eqnarray}
which is assume as the ground state of the spin model (\ref{eq:KAFH}) on the kagome lattice. This is the basic \emph{gound-state assumption}.

The ground GPWF $|G\rangle$ is a singlet state with $\mathbf{S}_{\text{tot}}=0$. Since we are interested in the spin dynamics, we need construct excited states with total $z$-component spin $S^z_{\text{tot}}=1$. Based on the mean field energy levels, we can construct many different states with total $S_{\text{tot}}^z=1$
\begin{eqnarray}
  \label{eq:excitedSfull}
  |S_{\text{tot}}^z=1\rangle=\mathcal{P}_G\left(\underbrace{e_{i_0}^\uparrow\cdots e_{i_{N/2}}^\uparrow}_{N/2+1}|\text{vac}\rangle\otimes\underbrace{e_{j_0}^\downarrow\cdots e_{j_{N/2-2}}^\downarrow}_{N/2-1}|\text{vac}\rangle\right)
\end{eqnarray}
where $e_{i_j}^\uparrow\in\{e_0^\uparrow,\cdots,e_{N-1}^\uparrow\}$ and $e_{j_i}^\downarrow\in\{e_0^\downarrow,\cdots,e_{N-1}^\downarrow\}$. The dimension of sub Hilbert space with $S^z_{\text{tot}}=1$ is
\begin{eqnarray}
\label{eq:dimHF}
  \text{Dim}(\mathcal{H}_{\text{full}}[S^z_{\text{tot}}=1])\sim\begin{pmatrix}N\\N/2+1\end{pmatrix}^2.
\end{eqnarray}

\emph{Excited-state assumption:} the second assumption is that we are only interested in excited states which can be constructed by spinon particle-hole excited states
\begin{eqnarray}
  \label{eq:gpwfE}
  |S_{\text{tot}}^z=1\rangle=\mathcal{P}_G(e_{i\uparrow}^\dag e_{j\downarrow}|\Psi\rangle)
\end{eqnarray}
where $|\Psi\rangle = e_0^\uparrow\cdots e_{N/2-1}^\uparrow|\text{vac}\rangle\otimes e_0^\downarrow\cdots e_{N/2-1}^\downarrow|\text{vac}\rangle$  and $e_{i\uparrow}\in\{e_{N/2}^\uparrow,\cdots,e_{N-1}^\uparrow\}$, $e_{j\downarrow}\in\{e_0^\downarrow,\cdots,e_{N/2-1}^\downarrow\}$. This truncated sub-Hilbert space has the dimension
\begin{eqnarray}
\label{eq:dimH}
  \text{Dim}(\mathcal{H}[S^z_{\text{tot}}=1])\sim N^2/4.
\end{eqnarray}
Note in Eqs. (\ref{eq:dimHF}) and (\ref{eq:dimH}) we use $\sim$ instead of $=$, because those states may be not linearly independent. The dimension in Eq. (\ref{eq:dimHF}) $\sim\mathcal{O}(4^N), N\rightarrow\infty$ is much larger than the full spin Hilbert space dimension $2^N$ when $N$ is large. On square lattice for uniform RVB state, only around half of the excited states in Eq. (\ref{eq:gpwfE}) are linearly independent and $\text{Dim}(\mathcal{H}[S^z_{\text{tot}}=1])\sim N^2/8$.

 \subsection{Dynamic spin structure factor}
At zero temperature, the spin-spin correlation is given as
\begin{eqnarray}
  \label{eq:spincorr}
  \langle T_\tau S_i^-(\tau)S_j^+(0)\rangle_0=\langle G|e^{\tau H} S_i^{-}e^{-\tau H}S_j^+|G\rangle.
\end{eqnarray}
Under the above two GPWF assumptions, we can expand the Hamiltonian in the truncated Hilbert space
\begin{eqnarray}
\label{eq:pHamEx}
  H=\sum_n E_n|E_n\rangle\langle E_n|,
\end{eqnarray}
where $|E_n\rangle$ is the normalized orthogonal basis. Then we have the dynamical spin structure factor
\begin{eqnarray}
  \label{eq:chi}
  S(\mathbf{q},\omega)=\sum_n\delta(\omega-(E_n-E_0))|\langle E_n|S_{\mathbf{q}}^+|G\rangle|^2,
\end{eqnarray}
where the translational symmetry is assumed.

\subsection{Projected Hamiltonian system in truncated sub-Hilbert space}
We are working on the truncated sub-Hilbert state with total spin $S^z_{\text{tot}}=1$ in Eq. (\ref{eq:gpwfE}). We should project the spin Hamiltonian onto the truncated Hilbert space to obtain the expansion as in Eq. (\ref{eq:pHamEx}).

Due to projection, the Gutzwiller-projected particle-hole states in Eq. (\ref{eq:gpwfE}) are not orthogonal any more (even not linearly independent). To obtain the normal orthogonal basis, we need calculate the overlaps between these states and the Hamiltonian matrix
\begin{eqnarray}
  \label{eq:overlap}
\mathbb{O}(i,j)=\langle i|j\rangle, \quad \mathbb{H}(i,j)=\langle i|H|j\rangle,
\end{eqnarray}
Diagonalize the overlap matrix
\begin{eqnarray}
  \mathbb{O}V^{(n)}=\lambda_n V^{(n)},
\end{eqnarray}
we obtain the normal orthogonal basis
\begin{eqnarray}
  |\alpha_n\rangle = \sum_{i=0}^{N^2/4-1}\frac{1}{\sqrt{\lambda_n}}|i\rangle V_i^{(n)}
\end{eqnarray}
Note here zero eigenvalues $\lambda_n$ imply the $|\alpha_n\rangle$ is not an independent state and should be removed from the orthogonal basis. The rank $r$ of overlap matrix $\mathcal{O}$ is the dimension of the truncated sub-Hilbert space, $\text{Dim}(\mathcal{H}[S^z_{\text{tot}}=1])=r$. Generally, $r<=N^2/4$, e.g., uniform RVB state on the square lattice has $r\sim N^2/8$.

Based on the orthogonal basis $|\alpha_n\rangle$, we can obtain the Hamiltonian
\begin{eqnarray}
  \langle\alpha_m|H|\alpha_n\rangle = V_i^{(m)*}\mathbb{H}(i,j)V_j^{(n)}/\sqrt{\lambda_m\lambda_n}.
\end{eqnarray}
Diagonalize the above Hamiltonian matrix, we can obtain the normal orthogonal expansion of Hamiltonian in Eq. (\ref{eq:pHamEx}). When the overlap matrix $\mathcal{O}$ has the full rank, the above procedure can be re-expressed in the generalized eigenvalue problem\cite{Li2010}
\begin{eqnarray}
  \label{eq:genEigen}
  \mathbb{H}\phi^{n}=E_n\mathbb{O}\phi^{n}
\end{eqnarray}
where $|E_n\rangle=\sum_{i=0}^{N^2/4-1}|i\rangle \phi_i^{(n)}$.
Therefore the dynamic spin susceptibility is given as
\begin{eqnarray}
  \label{eq:chiGen}
  \chi(q,\omega)=\sum_n\delta(\omega-(E_n-E_0))|\sum_{ij}\phi_i^{(n)*}\mathcal{O}(i,j)\varphi_j^S|^2,\nonumber\\
\end{eqnarray}
where the spin operator is expressed as $S_q^+=\sum_{i=0}^{N^2/4-1}|i\rangle \varphi_i^{S}$.

For reasonably large system size, e.g. $N = 12 \times 6 \times 3$ on the kagome lattice, the dimension $\text{Dim}(\mathcal{H}[S^z_{\text{tot}}=1])$ is still huge for numerical calculations and large memories are needed to store $\mathbb{O}$ and $\mathbb{H}$ matrices. If the system has the translational symmetry, we can decompose $\text{Dim}(\mathcal{H}[S^z_{\text{tot}}=1])$ according to the lattice momentum $\mathbf{q}$
\begin{eqnarray}
  \text{Dim}(\mathcal{H}[S^z_{\text{tot}}=1])=\bigoplus_{\mathbf{q}}\text{Dim}(\mathcal{H}[S^z_{\text{tot}}=1;\mathbf{q}]).
\end{eqnarray}
Particularly, for FSL state, $\text{Dim}(\mathcal{H}[S^z_{\text{tot}}=1;\mathbf{q}])$ are not the same for different momentum $\mathbf{q}$. The typical value of the dimension is $\text{Dim}(\mathcal{H}[S^z_{\text{tot}}=1;\mathbf{q}])\sim \mathcal{O}(N)$, which is small enough for the efficient numerical calculations.

\subsection{Monte Carlo algorithm}
The key evaluations are the matrix elements for $\mathbb{O}$ and $\mathbb{H}$ matrices in Eq. (\ref{eq:overlap}). This can be done by Monte Carlo method. We will take the sampling strategy developed by Li and Yang in Ref. \onlinecite{Li2010}: using \emph{single} Markov chain to generate spin configurations for all element evaluations:
\begin{eqnarray}
  \mathbb{O}(i,j)&=&\sum_{\alpha}\frac{\langle i|\alpha\rangle\langle\alpha|j\rangle}{\rho(\alpha)}\rho(\alpha),\nonumber\\
  \mathbb{H}(i,j)&=&\sum_{\alpha}\frac{\langle i|\alpha\rangle\langle H\alpha|j\rangle}{\rho(\alpha)}\rho(\alpha),
\end{eqnarray}
where $|\alpha\rangle$  is spatial spin configurations generated according to the Monte Carlo sampling probability $\rho(\alpha)$. Given the spatial spin configuration $|\alpha\rangle$, Gutzwiller-projected particle-hole states in Eq. (\ref{eq:gpwfE}) are all Slater determinants
\begin{eqnarray}
  \langle\alpha|i\rangle = \langle\alpha_\uparrow|i_\uparrow\rangle\times\langle\alpha_\downarrow|i_\downarrow\rangle.
\end{eqnarray}
For the evaluations, we would pick up one \emph{reference state} which is the lowest mean field particle-hole state
\begin{eqnarray}
  |R\rangle = \mathcal{P}_G\left(e_0^\uparrow\cdots e_{N/2}^\uparrow|\text{vac}\rangle\otimes e_0^\downarrow\cdots e_{N/2-2}^\downarrow|\text{vac}\rangle\right).
\end{eqnarray}
and calculate its Slater determinant
\begin{eqnarray}
  \langle\alpha|R\rangle = \langle\alpha_\uparrow|R_\uparrow\rangle\times\langle\alpha_\downarrow|R_\downarrow\rangle.
\end{eqnarray}
Since every determinant $\langle\alpha_{\uparrow/\downarrow}|i_{\uparrow/\downarrow}\rangle$ differs with $\langle\alpha_{\uparrow/\downarrow}|R_{\uparrow/\downarrow}\rangle$ only by one column, it is easy to evaluate using rank-1 determinant update. The computation complexity is $\mathcal{O}(N)$. The Hamiltonian matrix element determinant is
\begin{eqnarray}
  \langle H\alpha_{\uparrow/\downarrow}|i_{\uparrow/\downarrow}\rangle
\end{eqnarray}
which differs with  $\langle\alpha_{\uparrow/\downarrow}|R_{\uparrow/\downarrow}\rangle$ by one row and one column and also can be evaluated using rank-2 determinant update. The computation complexity is $\mathcal{O}(N^3)$. So the full problem has the complexity is $\mathcal{O}(N^3)$.

In Refs. \onlinecite{Li2010}, the total weight of all states with the same momentum $\mathbf{q}$ is used for $\rho_\mathbf{q}(\alpha)=\sum_i |\langle\alpha|i,\mathbf{q}\rangle|^2$. During the update according the total weight, the matrices for the reference state determinants $\langle\alpha_\uparrow|R_\uparrow\rangle $ or $\langle\alpha_\downarrow|R_\downarrow\rangle$ may be singular. To avoid the singularity, we use the probability function
\begin{eqnarray}
  \rho(\alpha)= |\langle\alpha_\uparrow|R_\uparrow\rangle\times\langle\alpha_\downarrow|R_\downarrow\rangle|^2.
\end{eqnarray}

\bibliography{SDkagome}